\documentclass[a4paper]{jpconf}
\usepackage{graphicx}
\usepackage{amsmath}
\usepackage{mathrsfs} 
\usepackage{amsfonts}
\usepackage{dsfont}
\usepackage{feynmp}
\usepackage{caption}

\begin{document}
\title{\(\tau\) vector and axial vector spectral functions in the extended linear sigma model}

\author{A. Habersetzer$^{1}$, F. Giacosa$^{1,2}$}

\address{$^{1}$Institut f{\"u}r Theoretische Physik, Goethe-Universit{\"at}, Max-von-Laue-Str. 1, 60438 Frankfurt, Germany \\
$^{2}$ Institute of Physics, Jan Kochanowski University, 25-406 Kielce, Poland}

\ead{habersetzer@th.physik.uni-frankfurt.de, fgiacosa@ujk.edu.pl}

\begin{abstract}
The extended linear sigma model describes the vacuum phenomenology of scalar, pseudoscalar, vector and axial-vector mesons at energies \(\simeq 1\text{ GeV}\). We combine the chiral \(U(2)_L\times U(2)_R\) symmetry of this model with a local \(SU(2)_L\times U(1)_Y\) symmetry and obtain a gauge invariant effective description for electroweak interaction of hadrons in the vacuum. Vector and axial-vector spectral functions can be described well by two intermediate resonances \(\rho\) and \(a_1\). They are implemented into this model as chiral partners and yield the predominant contributions to both spectral functions. However, the contributions that arise from the non-resonant decay channels of the weak charged \(W\) bosons are essential for reproducing the lineshapes of the spectral functions. \end{abstract}
  
\section{Introduction}
The weak decay of the \(\tau\) lepton as e.g. measured by the ALEPH collaboration \cite{Schael:2005am} provides an excellent ground for studying electroweak interactions at low energies. As a phenomenological approach vector meson dominance describes the interaction between the photon and hadrons by \(\bar{q} q\) vacuum fluctuations of the photon that form a neutral \(\rho\) meson which then interacts with other hadrons (e.g. pions). We can also understand weak interaction with hadrons in terms of such vacuum fluctuations that manifest themselves as mixing of the charged weak currents with vector and axial-vector mesons. The question whether the axial vector resonance \(a_1(1260)\) can indeed be described as a \(\bar{q}q\) state within the quark model, where it is the \(J^{PC}=1^{++}\) member of the \(N_F=2\) axial-vector multiplet, or whether it might not be a dynamically generated \(\rho\)-\(\pi\) molecule-like state has not been unambiguously answered, yet \cite{Wagner:2008gz,Roca:2006tr}. Related to this is also the question whether chiral models can describe the \(a_1\) and \(\rho\) resonance as chiral partners. Including weak interaction is thus not only useful to describe electroweak interactions of hadrons, it also provides a basis for understanding more about the nature of the \(a_1(1260)\) resonance and the applicability of an effective chiral model to the vacuum phenomenology. 

Effective chiral models (such as e.g. \cite{Gasiorowicz:1969kn}) that apply the principles of QCD to hadronic ob\-ser\-vables are widely used to study the low-energy phenomenology in heavy-ion collisions. In the extended Linear Sigma Model (eL\(\sigma\)M, see Refs. \cite{Parganlija:2012fy,Parganlija:2010fz,Janowski:2014ppa,Gallas:2009qp}) that we present here scalar, pseudoscalar, vector, and axial-vector fields are considered genuine degrees of freedom that can be constructed from a \(\bar{q}q\) picture. We show that with the eL\(\sigma\)M we can indeed describe the \(\tau\) vector and axial-vector resonances as emerging from the mixing of the chiral partners \(a_1\) and \(\rho\) with the charged weak \(W\) boson.

\section{Lagrangian}
The hadronic degrees of freedom are described by the matrix-valued fields 
\begin{align}
\Phi = (S_a+i P_a) t_a\ \!, \ L^\mu = (V_a^\mu + A^\mu_a) t_a \ \!, \ R^\mu_a = (V_a^\mu - A^\mu_a) t_a
\end{align}
on the basis of the strong isospin algebra with generators \(t_a=(1/2)\sigma_a\). 
Their masses, strong and weak interactions are obtained from the Lagrangian
{\small \begin{align}
  \begin{split} 
\mathscr{L} &= \text{Tr} \bigl[(D^\mu \Phi)^\dagger (D^\mu \Phi)\bigr] - m_0^2 \text{Tr}[\Phi^\dagger \Phi]  - \lambda_1\text{Tr} [(\Phi^\dagger \Phi)]^2 - \lambda_2 \text{Tr}[(\Phi^\dagger \Phi)^2]  - \frac{1}{4}\text{Tr}[(L^{\mu\nu})^2 + (R^{\mu\nu})^2] \label{eq:FullLagrangian} \\[3pt]
&   + \frac{m_1}{2}\text{Tr}[\left(L^{\mu\ \!\!2}\! + i \frac{g_2}{2} \bigl( \text{Tr} \bigl[ L_{\mu \nu} [ L_{\mu},L_{\nu} ] \bigr]   + \text{Tr} \bigl[R_{\mu \nu}[ R_{\mu},R_{\nu}] \bigr] \bigr)  + \! R^{\mu\ \!\!2}\right)] +\text{Tr}[H(\Phi + \Phi^\dagger)] \\
& + c_1 (\text{det}\Phi - \text{det}\Phi^\dagger)^2  + 2 h_3 \text{Tr}[\Phi R_\mu \Phi^\dagger L^\mu] + \frac{g \delta_w}{2} \text{Tr}[W_{\mu\nu}L^{\mu\nu}] + \frac{g' \delta_\text{em}}{2} \text{Tr}[R_{\mu\nu}B^{\mu\nu}] \\
&   +\frac{1}{4} \text{Tr}[(W^{\mu\nu})^2 +(  B^{\mu\nu})^2] +\frac{g}{2\sqrt{2}}\left( W_\mu^-\bar{u}_{\nu_\tau} \gamma_\mu (1-\gamma_5) u_\tau+\text{h.c.}\right)  + \mathscr{L}_{\Phi L R} + \mathscr{L}_{L R}  \ \!. 
\end{split} 
\end{align}} The Lagrangian is invariant under a {\bf global} linear chiral \(U(2)_L\times U(2)_R\) transformation 
\begin{align}
\Phi \rightarrow & U_L \Phi U_R^\dagger\ \!,\ L^\mu \rightarrow U_L L^\mu U_L^\dagger\ \!, \ R^\mu \rightarrow U_R R^\mu U_R^\dagger \ \!,
\end{align}
and it is also invariant under a {\bf local} \(SU(2)_L\times U(1)_Y\) transformation by
\begin{align}
\Phi \rightarrow & U_L \Phi U_Y^\dagger \ \!,\ L^\mu \rightarrow U_L L^\mu U_L^\dagger\ \!, \ R^\mu \rightarrow U_Y R^\mu U_Y^\dagger.
\end{align}
The bare gauge fields \(W^\mu = W_i^\mu t_i\) and \(B^\mu = B^\mu t_3\) transform in the adjoint representation of \(SU(2)_L\times U(1)_Y\) as
\begin{align}
W^\mu \ \xrightarrow{SU(2)_L} \  U_L W^\mu U_L^\dagger + \frac{i}{g} U_L \partial^\mu U_L^\dagger \ \!, \ B^\mu \ \xrightarrow{\ U(1)_Y\ \!} \  U_Y B^\mu U_Y^\dagger + \frac{i}{g'} U_Y \partial^\mu U_Y^\dagger \ .
\end{align}
After rotating the bare fields into the physical fields \(A_\mu\) and \(Z_\mu\) by the Weinberg mixing angle \(\theta_W\), we obtain the covariant derivative 
\begin{align}
 D^\mu \Phi  = & \ \ \partial^\mu \Phi - i g_1 (L^\mu \Phi - \Phi R^\mu) - i g \cos{\theta_C} (W^\mu_1 t_1 + W^\mu_2 t_2) \Phi - i e[A^\mu t_3,\Phi] \nonumber \\
& \!\!- i g \cos{\theta_W} (Z^\mu t_3 \Phi + \tan{\theta_W}\Phi Z^\mu t_3)\ \!. \label{eq:CovDer}
\end{align}
The Cabibbo angle \(\theta_C\) in (\ref{eq:CovDer}) rotates the flavor eigenstates into the weak eigenstates. The global chiral symmetry is broken spontaneously by the scalar condensate and explicitly by the term \(\sim H\) which modulates non-vanishing but equal current quark masses and also by the 't Hooft determinant that corresponds to the explicitly broken \(U(1)_A\) symmetry.
In (\ref{eq:FullLagrangian}) we included two terms \(\sim \delta_w\) and \(\sim \delta_{em}\). These terms describe a mixing between the electroweak interaction fields and the (axial-)vector fields. This mixing arises from the hadronic loop contributions of the electroweak bosons and is also described by vector meson dominance in its {\it first representation} as explained in \cite{O'Connell:1995wf}. The \(W\)-\(\rho\) mixing is only given by \(\sim \delta_w s\) while the \(W\)-\(a_1\) mixing contains an additional contribution \(\sim (g_1 \phi^2-\delta_w s)\) by the chiral condensate. 

\section{\protect\(\tau\protect\)-decay and spectral functions}
\begin{figure}[ht]\vspace{-2cm}
\includegraphics[width=1\textwidth]{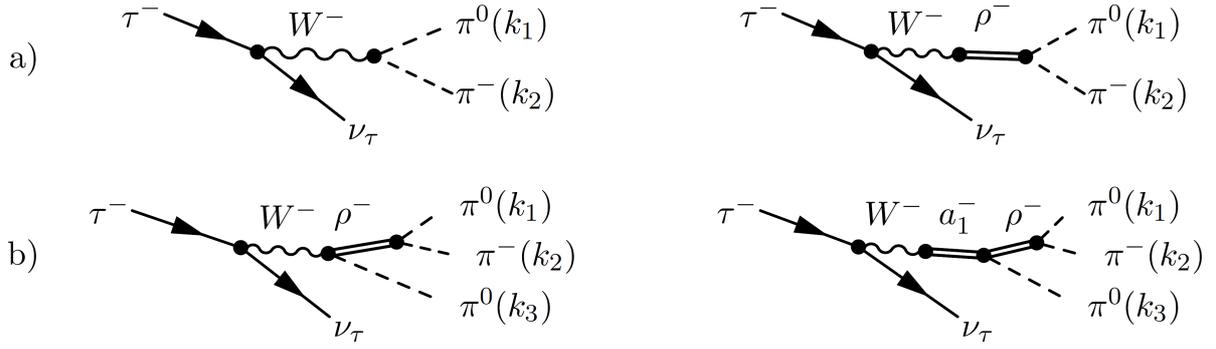} 
\caption{Feynman diagrams of a) the \(2 \pi\) decays in the vector channel and b) the \(3\pi\) decays of the axial-vector channel.}
\label{fig:FeynmanDiagramsVectorChannel} \end{figure}

In the vector channel \(\tau^-\) decays predominantly into \(\pi^-\pi^0 \nu_\tau\) and in the axial-vector channel pre\-domi\-nant\-ly by two isospin channels into \(2 \pi^0 \pi^-\) and \(2\pi^-\pi^+\). The Feynman diagrams we use are shown in Fig. \ref{fig:FeynmanDiagramsVectorChannel}. Because of the identical particles in the final state of the axial-vector channel we have to integrate over \(m_{12}^2\) and \(m_{23}^2\) and include the corresponding symmetry factor \(1/N! = 1/2\). In principle there are also the contributions from the direct \(W\rightarrow 3\pi\) and \(W\rightarrow a_1 \rightarrow 3\pi\) decays, but they are expected to be small and not included, yet. 
Decay rates and spectral functions are calculated on tree-level. We use the optical theorem \cite{Giacosa:2007bn} that relates the tree level amplitude to the imaginary part of the self-energy \(\text{Im}[\Sigma(s)] = \sqrt{s} \Gamma (s)\) and define vector and axial-vector propagator as
\begin{align}
D^{\mu\nu}_\rho(s) = \frac{-g_{\mu\nu}+\frac{P_\mu P_\nu}{P\cdot P}}{s-m_\rho^2 + i \sqrt{s}\ \! \Gamma_{\rho}(s)} \ \!, \quad D_{a_1}^{\mu\nu}(s) = \frac{-g_{\mu\nu}+\frac{P_\mu P_\nu}{P\cdot P}}{s-m_{a_1}^2 +i \sqrt{s}\ \! \Gamma_{a_1}(s)}\ \!.
\end{align} 
The bare masses are given by 
\begin{align}
m_\rho^2  = m_1^2 + \phi^2(h_1 + h_2 + h_3)/2\ \!,\ m_{a_1}^2  = m_1^2 + g_1^2 \phi^2 + \phi^2(h_1 + h_2 - h_3)/2 \ \!.
\end{align} 
They are fixed to the physical masses \(m_\rho\) and \(m_{a_1}\). We require that the real part of the respective self-energies \(\text{Re}\Sigma(s=m^2)=0\). Then the contributions that arise from the real part of the self-energies will not contribute significantly to the spectral functions in the region around the peak.
The sum rules
\begin{align}
\int\limits_0^\infty ds \ \! d_{\rho}(s) = 1 \ \!, \ \int\limits_0^\infty ds \ \! d_{a_1}(s) = 1 \label{eq:sumrule}
\end{align}
describe a probability conservation where \(d(s)\) is the probability that \(\rho\) and/or \(a_1\) meson have mass \(m=\sqrt{s}\). The sum rule in (\ref{eq:sumrule}) holds only at all orders in resummed perturbation theory. Care is needed when non-renormalizable interaction terms are considered and if moreover, \(\text{Im}\Sigma(s)\) is only calculated at one loop with its real part being neglected. Therefore we need to introduce a `cutoff', which is usually of the order of \(1.5-2\text{ GeV}\). Here we choose \(m_\tau\) and obtain the normalized spectral functions as
\begin{align}
\frac{1}{N_\rho} \int\limits_0^{m_\tau^2} ds \ \! d_{\rho^- \rightarrow \pi^-\pi^0}(s) = 1 \ \!, \quad  \frac{1}{N_{a_1}} \int\limits_0^{m_\tau^2} ds \ \! d_{a1}(s) = 1\ \!.
\end{align}
We can then obtain vector and axial-vector spectral functions from
\begin{align}
d_{V}(s) = \frac{1}{\pi} \frac{\sqrt{s} \ \Gamma_{W^-\rightarrow \pi^-\pi^0}(s)}{M_w^4}\ \!, \quad d_{A}(s) = \frac{1}{\pi} \frac{\sqrt{s} \ \Gamma_{W^-\rightarrow3\pi}(s)}{M_w^4}\ \!,
\end{align}
where \(\Gamma_W(s)\) contains the coherent amplitude squared of the processes depicted in Fig. \ref{fig:FeynmanDiagramsVectorChannel} and where, for \(s \ll M_w^2\), the \(W\) propagator reduces to a point-like interaction vertex. Notice that, if the direct contributions \(W^-\rightarrow \pi^0\pi^-\) and \(W^- \rightarrow\rho^{0/-}\pi^{-/0}\) were neglected we would obtain \(d_V(s)=d_{W\rightarrow\rho\rightarrow 2\pi}(s)\) and \(d_A(s)=d_{W\rightarrow a_1\rightarrow 3\pi}(s)\). But, as we will see in the next section, this limit yields only an approximate description of the spectral functions.

\section{Results}
All relevant parameters, that is \(g_1\ \!,\ g_2\ \!,\ h_3 \ \!,\ w\), and \(Z\), except the mixing parameter \(\delta_w\), have been determined in a global fit in \cite{Parganlija:2012fy}. We use their result for masses and decay widths
\begin{align}
&m_\rho = (0.7831 \pm 0.0070) \text{ GeV}\ \!,& \ &\Gamma_\rho=(0.1609 \pm 0.0044)\text{ GeV} \ \!, \\
&m_{a_1} = (1.186 \pm 0.006) \text{GeV}\ \!,& \ &\Gamma_{a_1} =(0.549 \pm 0.043 )\text{GeV}\ \! ,
\end{align}
such that \(\delta_w\) is the only free paramter. The result for the coherent vector-channel spectral function compared to the ALEPH data is seen in Fig. \ref{fig:01VectorchannelCoherent}. 

\vspace{0.4cm}\hspace{-0.5cm}\begin{minipage}[t][0.36\textheight][t]{0.49\textwidth}
\includegraphics[width=1\textwidth]{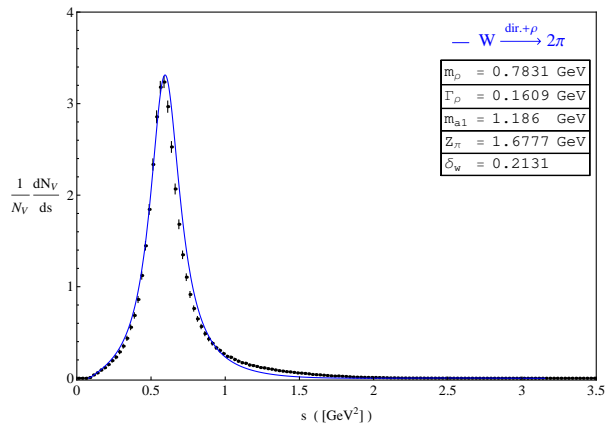} 
\captionof{figure}{Vector channel spectral function of the coherent sum of the diagrams in Fig. \ref{fig:FeynmanDiagramsVectorChannel}a).} 
\label{fig:01VectorchannelCoherent} 
\end{minipage}\hspace{0.3cm}
\begin{minipage}[t][0.36\textheight][t]{0.49\textwidth}
\includegraphics[width=1\textwidth]{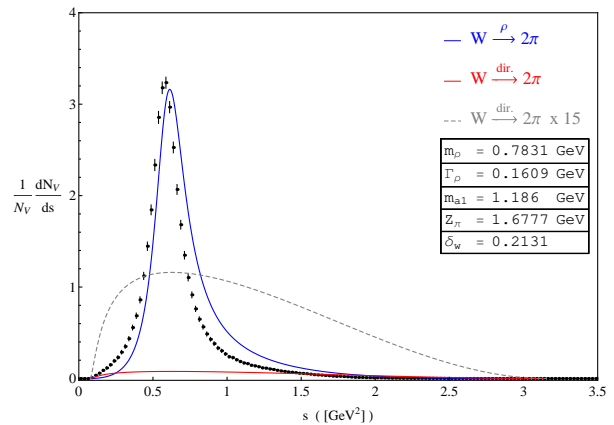} 
\captionof{figure}{Individual contributions; the blue line represents the \(W\)-\(\rho\) mixing, the red line \(W^-\rightarrow \pi^- \pi^0\), and the dashed gray line the direct contribution scaled by \(15\).} 
\label{fig:02VectorChannelIndContributions} 
\end{minipage}
The coupling \(\delta_w\) that describes the \(W\)-\(\rho\) mixing is determined by the peak value of the mass distribution \((1/N_{V}) dN_{V}/ds\). Apart from a small shift of the peak to higher energies the spectral function nicely describes the experimental distribution. This shift is expected, since if we look at the results of the global fit in \cite{Parganlija:2012fy} we see that also there \(\Gamma_{\rho^- \rightarrow \pi^-\pi^0}\) deviates slightly from the experimental data. The individual contributions to the coherent sum are shown in Fig. \ref{fig:02VectorChannelIndContributions}. The process \(\tau^- \rightarrow \pi^-\pi^0\) is clearly dominated by the intermediate \(\rho\) meson which constitutes almost the entire strength of the vector spectral function. However, the numerically small contribution of the direct \(W^- \rightarrow \pi^-\pi^0\) decay has a strong influence on the lineshape, particularly visible e.g. at \(s>m_\rho^2\), where due to destructive quantum interference the excess that arises from the \(\rho\) resonance contribution disappears. This can be interpreted as a very nice demonstration of VMD. In the low energy region the charged weak interaction fields seem to essentially couple to the pions by forming a vector meson resonance. 

The coherent sum in the axial-vector channel shown in Fig. \ref{fig:03AxialvectorCoherent} presents a similar picture. While the \(a_1\) meson yields the strongest contribution, the influence of the direct decay \(W^-\rightarrow \rho\pi\) is clearly visible. Notice that we have not yet included the direct contributions from \(W\rightarrow 3\pi\) and \(a_1\rightarrow 3\pi\). This is probably the reason why the axial-vector spectral function lacks some strength at the peak and above \(2\text{ GeV}^2\). In the non-resonant \(a_1\rightarrow 3\pi\) decay there are however unknown parameters that have yet to be determined. Thus, including these contributions would not necessarily render the result more conclusive. We could however use the ALEPH data also to obtain a first estimate on these unknown parameters. It should be stressed once more, that for both spectral-functions we have only one free parameter, the mixing constant \(\delta_w\), which is restrained by the peak value of the vector channel spectral function, while all other parameters are taken from the global fit in \cite{Parganlija:2012fy}.

\vspace{0.3cm}\hspace{-0.5cm}\begin{minipage}[t][0.34\textheight][t]{0.49\textwidth}
\includegraphics[width=1\textwidth]{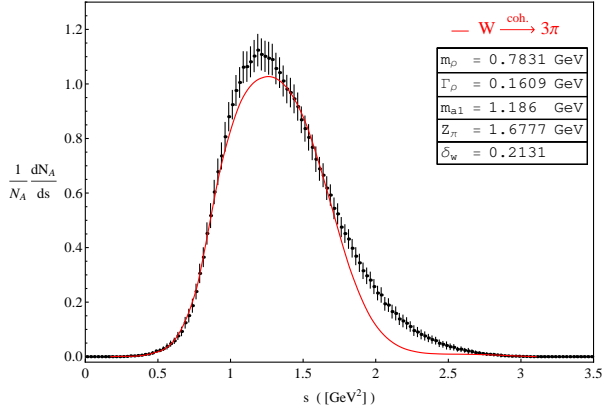} 
\captionof{figure}{ The axial-vector spectral function obtained from the coherent sum of the diagrams in Fig. \ref{fig:FeynmanDiagramsVectorChannel} b).} 
\label{fig:03AxialvectorCoherent} 
\end{minipage}\hspace{0.3cm}
\begin{minipage}[t][0.34\textheight][t]{0.49\textwidth}
\includegraphics[width=1\textwidth]{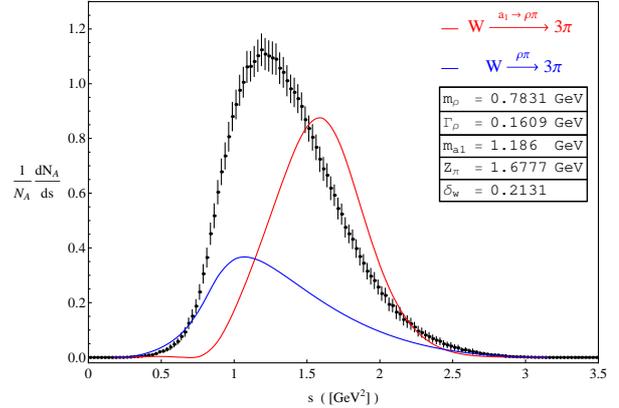} 
\captionof{figure}{Individual contributions to the axial-vector spectral function. The blue lines represents \(W \rightarrow a_1\rightarrow \rho \pi \) and the red line \(W \rightarrow a_1\rightarrow \rho \pi \).} 
\label{fig:04AxialvectorIndividualContributions} 
\end{minipage} Within this fit all parameters have been determined to \(\pm 5\%\) accuracy. We can still adjust the parameters within this range and obtain an even better description.

\section{Conclusion}
Together with a gauge invariant inclusion of an external electroweak field the eL\(\sigma\)M can describe vector and axial-vector channels of the electroweak decay of the \(\tau\). The difference in the coupling of the weak bosons to \(\rho\) and \(a_1\) arises only from the chiral condensate because of the spontaneously broken chiral symmetry. In our chiral framework the axial-vector resonance can indeed be described as a \(\bar{q}q\) meson and as the chiral partner of \(\rho\). In both channels we also obtained a nice demonstration of vector meson dominance. Although, to some extent, the \(\tau\) decay can be described by \(\rho\) and \(a_1\) resonances only, the contributions from \(W\) decay are clearly needed. These results can still be improved by adjusting the parameters to this specific process. Moreover, we can still include the direct decays \(W\rightarrow 3\pi\) and \(a_1\rightarrow 3\pi\) into the coherent sum and see how they will influence the spectral function.

\ack The authors acknowledge support from the Helmholtz Research School on Quark Matter
Studies and thank D. H. Rischke for his collaboration and also M. Buballa, J. Wambach, H. v. Hees, G. Wolf, and Z. Szep for valuable discussions.


\section*{References}
\begin{thebibliography}{99}
\bibitem{Schael:2005am}
  Schael S et al. 2005
  {\it Phys.\ Rept.}\  {\bf 421} 191

\bibitem{Wagner:2008gz}
  Wagner M and Leupold S 2008
  {\it Phys.\ Rev.}\ D {\bf 78}  053001
	
\bibitem{Roca:2006tr}
  Roca L, Oset E and Singh J 2006
  {\it AIP Conf.\ Proc.}\  {\bf 814} 468.
	

\bibitem{Gasiorowicz:1969kn}
  Gasiorowicz S and Geffen D A 1969
 {\it Rev.\ Mod.\ Phys.}\  {\bf 41} 531.

\bibitem{Parganlija:2012fy}
  Parganlija D, Kovacs P, Wolf G, Giacosa F and Rischke D H 2012
  {\it Phys.\ Rev.}\ D {\bf 87} 014011

\bibitem{Parganlija:2010fz}
  Parganlija D, Giacosa F and Rischke D H 2010 
  {\it Phys.\ Rev.\ } D {\bf 82} 054024
	
\bibitem{Janowski:2014ppa}
  Janowski S, Giacosa F and Rischke D H 2014
  arXiv:1408.4921 [hep-ph].
	
\bibitem{Gallas:2009qp}
  Gallas S, Giacosa F and Rischke D H 2010
  {\it Phys.\ Rev.\ } D {\bf 82} 014004
	
\bibitem{O'Connell:1995wf}
  O'Connell H B, Pearce C, Thomas A W and Williams A G 1997
  {\it Prog.\ Part.\ Nucl.\ Phys.\ }  {\bf 39} 201

	
\bibitem{Giacosa:2007bn}
  Giacosa F and Pagliara G 2007  
  {\it Phys.\ Rev.\ } C {\bf 76} 065204
\end{thebibliography}
\end{document}